\documentclass[aps,pra,reprint,superscriptaddress]{revtex4-2}

\usepackage{amsmath,amsfonts,amssymb}
\usepackage{xcolor,graphicx}
\usepackage{braket}
\usepackage[pdftex]{hyperref}
\hypersetup{
	colorlinks = true,
	linkcolor = blue,
	anchorcolor = blue,
	citecolor = red,
	filecolor = blue,
	urlcolor = blue}

\begin{document}

\title{Poynting vector for Cauchy-Riemann beams}

\author{I. Juli\'an-Mac\'ias}
\email[e-mail: ]{isjuma@inaoep.mx}
\affiliation{Instituto Nacional de Astrofísica Óptica y Electrónica, Calle Luis Enrique Erro No. 1\\ Santa María Tonantzintla, Puebla, 72840, Mexico}

\author{F. Soto-Eguibar}
\affiliation{Instituto Nacional de Astrofísica Óptica y Electrónica, Calle Luis Enrique Erro No. 1\\ Santa María Tonantzintla, Puebla, 72840, Mexico}

\author{I. Ramos Prieto}
\affiliation{Instituto Nacional de Astrofísica Óptica y Electrónica, Calle Luis Enrique Erro No. 1\\ Santa María Tonantzintla, Puebla, 72840, Mexico}

\author{U. Ruiz}
\affiliation{Instituto Nacional de Astrofísica Óptica y Electrónica, Calle Luis Enrique Erro No. 1\\ Santa María Tonantzintla, Puebla, 72840, Mexico}

\author{N. Korneev}
\affiliation{Instituto Nacional de Astrofísica Óptica y Electrónica, Calle Luis Enrique Erro No. 1\\ Santa María Tonantzintla, Puebla, 72840, Mexico}

\author{D. {S\'anchez}-{de-la-Llave}}
\affiliation{Instituto Nacional de Astrofísica Óptica y Electrónica, Calle Luis Enrique Erro No. 1\\ Santa María Tonantzintla, Puebla, 72840, Mexico}

\author{H. M. Moya-Cessa}
\affiliation{Instituto Nacional de Astrofísica Óptica y Electrónica, Calle Luis Enrique Erro No. 1\\ Santa María Tonantzintla, Puebla, 72840, Mexico}

\date{\today}

\begin{abstract}
    We present a detailed derivation of the Poynting vector for Cauchy-Riemann beams propagating in free space considering a Gaussian modulation with $g \in \mathbb{C}$. The effect generated by this Gaussian modulation is a compression-expansion of the intensity distribution. It is shown that the parameter $g$ can  reverse the direction of energy flux and eliminate the radial component, resulting in a purely azimuthal field. Additionally, we validate our analytical results through experimental verification.
\end{abstract}
\maketitle

The derivation of solutions to the paraxial wave equation typically involves exploring different coordinate systems, applying the Fresnel approximation, or utilizing quantum mechanical techniques. The most straightforward and widely examined case is light propagation in free space. Numerous solutions have been proposed, including Hermite-Gauss~\cite{BoydHermite}, Bessel-Gauss\,\cite{GoriBesselGauss}, Laguerre-Gauss\,\cite{AllenOrbital}, Ince-Gauss\,\cite{BandresInce}, Airy\,\cite{SiviloglouAiry}, Helmholtz-Gauss\,\cite{GutierrezHelmholtz}, Whittaker-Gauss\,\cite{MagoPropagation}, stable caustics\,\cite{EspindolaParaxial}, Durnin-Whitney-Gauss\,\cite{JulianThevector}, Tricomi-Gauss\,\cite{SinghTricomiGauss}, and Cauchy-Riemann\,\cite{MoyaCauchy} beams. Among these, some are notable for their ability to transfer orbital angular momentum, such as Bessel-Gauss\,\cite{VolkeThreedimensional} and Laguerre-Gauss\,\cite{PadgetThePoynitng,AllenIVTheOrbital} beams. Additionally, the perfect vortex beam is recognized for its angular momentum transfer capabilities\,\cite{OstrovskyLiquid,ChenDynamics}. Among these solutions, Cauchy-Riemann beams that propagate in both free space\,\cite{MoyaCauchy} and gradient-index media\,\cite{IranCauchyGRIN} are derived by solving the paraxial wave equation using quantum mechanical methods\,\cite{Stoler,KorneevUnified}. These beams are termed Cauchy-Riemann because the optical field at $z = 0$ is expressed as the product of a Gaussian function and an entire function $f(x+iy)$, which satisfies the Cauchy-Riemann conditions (for details on asymmetric Gaussian modulation see\,\cite{KorneevAsymmetric}). A distinctive characteristic of these paraxial beams is the rotation of their intensity distribution during propagation; however, the case of $g \in \mathbb{C}$ has not been explored, nor have the consequences for the Poynting vector. Addressing these two fundamental aspects is one of the primary objectives of this letter. It is important to mention that, some paraxial optical fields whose intensity distribution rotates during its propagation  have been reported previously\,\cite{SchechnerWave, BekshaevAnoptical, RazuevaMultiple}.

In this letter, we investigate Cauchy-Riemann beams in free space, where the optical field at $z = 0$ is generally described by an entire function $f(x + iy)$ combined with Gaussian modulation. We focus on two cases: (a) $f_1(x + iy) = (x + iy)^n$ and (b) $f_2(x + iy) = 1 - (x + iy)^n$, with $n \in \mathbb{N}$. The first case represents a field with topological charge determined by $n$, while in the second case the polynomial $1 - (x+iy)^n$ is closely related to cyclotomic polynomials which describe the $n$-th roots of unity in the complex plane\,\cite{Dummit}. We derive analytical expressions for intensity, phase distributions, and energy flux, and present detailed formulations for the Poynting vector, accompanied by graphical representations of intensity, phase, and transverse Poynting vector at various $z$ planes. Additionally, we analyze how the real and imaginary components of $g$ influence the flux direction and azimuthal behavior. The results are supported by experimental field synthesis.

\textit{Poynting vector in free space.} Recently, a new solution to the paraxial wave equation in free space has been introduced, considering a Gaussian modulation and an entire function\,\cite{MoyaCauchy}:
\begin{equation} 
\psi(\mathbf{r}_\perp,z) = \frac{\exp\left[-\dfrac{g(x^2+y^2)}{\omega(z)}\right]}{\omega(z)} f\left(\frac{x + iy}{\omega(z)}\right),\label{psiCR} 
\end{equation}
where $\mathbf{r}_\perp = (x, y)$, $\omega(z) = 1 + i2gz/k$ (with $k = 2\pi/\lambda$ being the wave number), $g \in \mathbb{C}$, and $f$ must satisfy the Cauchy-Riemann conditions. Importantly, since the set of complex functions $f$ that meet the Cauchy-Riemann conditions is extensive, there is a wide class of Cauchy-Riemann beams, each one characterized by different choices of $f$.

To investigate the structure of the Poynting vector for these beams, it is useful to consider $g = g_\mathrm{R} + i g_\mathrm{I}$ and $\omega(z) = \omega_\mathrm{R} + i \omega_\mathrm{I}$, where the subscripts $\mathrm{R}$ and $\mathrm{I}$ denote the real and imaginary parts of the complex numbers. After performing some algebraic manipulations in cilyndrical circular coordinates~$(r,\,\theta,\,z)$, we find that the Cauchy-Riemann beam described by Eq.~(\ref{psiCR}) can be expressed in the following form:
\begin{equation}\label{CRB1}
    \begin{split}
    \psi(r,\theta,z) & =  R(r,\theta,z)\exp\left[i\chi(r,\theta,z)\right],\\
    R(r,\theta,z)    & =  \frac{\exp\left(\dfrac{-r^2}{r_\omega ^2}A\right)}{r_\omega}\sqrt{u^2(r,\theta,z) + v^2(r,\theta,z)},\\
    \chi(r,\theta,z) & =  \Phi(r,\theta,z) + \displaystyle\frac{r^2}{r_\omega ^2} B - \theta_\omega,
    \end{split}
\end{equation}
where $A = g_R \omega_R + g_I \omega_I$, $B = g_R \omega_I - g_I \omega_R$, $r = \sqrt{x^2+y^2}$, $r_\omega = \sqrt{\omega_R^2 + \omega_I ^2}$, $\theta = \arg(x+iy)$ and $\theta_\omega = \arg[\omega(z)]$. Moreover, $u(r,\theta,z)$ and $v(r,\theta,z)$ are the real and imaginary parts of $f(\frac{r}{r_\omega}e^{i(\theta-\theta_\omega)})$, respectively, while $\Phi(r,\theta,z)=\arctan\left(v/u\right)$.

\begin{figure}
    \centering
    \includegraphics[width = \linewidth]{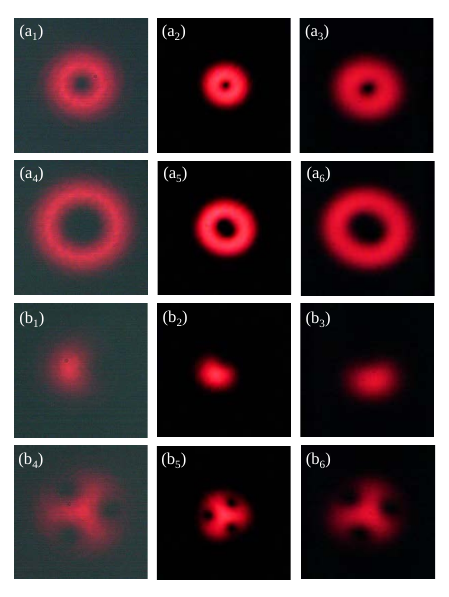}
    \caption{Experimental intensity distribution ($\mathrm{a}_1$)-($\mathrm{a}_3$) and ($\mathrm{a}_4$)-($\mathrm{a}_6$) determined by Eq.\,(\ref{psiCR}) considering $f_1\left(\frac{r}{r_\omega} e^{i(\theta-\theta_\omega)}\right) = \left(\frac{\alpha r}{r_\omega} e^{i(\theta-\theta_\omega)}\right)^n$, at the planes $z = 0\,\mathrm{m},\,z_0,\,2z_0$ for $n = 1$ and $n = 3$, respectively. Experimental intensity distribution ($\mathrm{b}_1$)-($\mathrm{b}_3$) and ($\mathrm{b}_4$)-($\mathrm{b}_6$) determined by Eq.\,(\ref{psiCR}) considering $f_2\left(\frac{r}{r_\omega} e^{i(\theta-\theta_\omega)}\right) = 1 - \left(\frac{\alpha r}{r_\omega} e^{i(\theta-\theta_\omega)}\right)^n$, at the planes $z = 0\,\mathrm{m},\,z_0,\,2z_0$ for $n = 1$ and $n = 3$, respectively. In these plots $g_R = .8 \times 10^{6}\,\mathrm{m}^{-2}$, $g_I = 1.5 \times 10^{6}\,\mathrm{m}^{-2}$, $\alpha = 8 \times 10^{4}\,\mathrm{m}$ and $\lambda = 633\,\mathrm{nm}$. The $x$ and $y$ coordinates ranging from  $-3~\mathrm{mm}$ to $3~\mathrm{mm}$.}
    \label{fig_1}
\end{figure}

Once Eq.\,(\ref{CRB1}) has been established, the Poynting vector associated with a Cauchy-Riemann beam is given by the following expression\,\cite{BerryExact}:
\begin{equation}
    \begin{split}
    \mathbf{J}(r,\theta,z) &= R^2(r, \theta,z)\bigg[  \hat{\mathbf{r}} \left(\frac{\partial \Phi(r,\theta,z)}{\partial r}  + \displaystyle\frac{2r}{r_\omega ^2} B \right) \\&+ \frac{\hat{\boldsymbol{\theta}}}{r}\frac{\partial \Phi(r,\theta,z)}{\partial \theta} + \hat{\mathbf{z}}\bigg].
    \end{split}
\end{equation}
As was mentioned, the set of entire functions is extensive; the most used example in literature, on entire functions, are the polynomial functions. In this letter, we consider that the function $f$ in Eq.\,(\ref{psiCR}) takes the following expressions: $[(x+iy)/\omega(z)]^n$ and $1-[(x+iy)/\omega(z)]^n$. Due to the expressions of the entire functions considered, it is convenient to express them in cylindrical coordinates, were $x+iy = re^{i\theta}$ and $z(\omega) = r_\omega e^{i\theta_\omega}$, so the entire functions can be expressed in the form 
\begin{eqnarray}
    f_1\left(\frac{r}{r_\omega} e^{i(\theta-\theta_\omega)}\right)  & = & \left(\frac{r}{r_\omega} e^{i(\theta-\theta_\omega)}\right)^n,\label{f_1}\\
    f_2\left(\frac{r}{r_\omega} e^{i(\theta-\theta_\omega)}\right) & = & 1 - \left(\frac{r}{r_\omega} e^{i(\theta-\theta_\omega)}\right)^n.\label{f_2}
\end{eqnarray}
Therefore, from the relations in Eq.\,(\ref{CRB1}), the amplitude and phase for these two examples are determined by
\begin{equation}\label{Rschis}
    \begin{split}
    R_1(r,\theta,z) = & \frac{\exp\left[\dfrac{-r^2}{r_\omega ^2}  A\right] r^n}{r_\omega^{n+1}},\\
    \chi_1(r,\theta, z) = & \frac{r^2}{r_\omega ^2} B + n\theta - (n+1)\theta_\omega,\\
    R_2(r,\theta,z) = & \frac{\exp\left[\dfrac{-r^2}{r_\omega ^2} A \right]}{r_\omega ^{n+1}} \sqrt{r_\omega ^{2n} - 2r_\omega^n r^n \cos\Omega + r^{2n}},\\
\chi_2(r,\theta,z)  = & \frac{r^2}{r_\omega ^2} B + \arctan \left(\dfrac{-r^n\sin\Omega}{r_\omega ^n - r^n \cos\Omega} \right) - \theta_\omega,
\end{split}
\end{equation}
where $ \Omega = n(\theta-\theta_\omega)$, and their Poynting vectors can be expressed as
\begin{equation}\label{Js}
    \begin{split}
        \mathbf{J}_1(r,\theta,z)&=  \frac{r^{2n-1} \exp\left[\dfrac{-2r^2}{r_\omega ^2 } A \right]}{r_\omega ^{2n+4}} \left[ 2r^2 B \hat{\mathbf{r}} + n r_\omega^2 \hat{\boldsymbol{\theta}}+\hat{\mathbf{z}} \right],\\
        \mathbf{J}_2(r,\theta,z) &=  \frac{\exp\left[\dfrac{-2r^2}{r_\omega ^2} A\right]}{r_\omega ^{2n+2}}\\
        &\times \left\{ n r^{n-1} \left[ - r_\omega ^n \sin\Omega \hat{\mathbf{r}} + (r^n-r_\omega^n \cos\Omega) \hat{\boldsymbol{\theta}}) \right] + \Delta \hat{\mathbf{z}} \right\},
    \end{split}
\end{equation}
where $\Delta = (r_\omega ^{2n} - 2r_\omega^n r^n \cos\Omega + r^{2n})$. Note that the radial component of the Poynting vector  $\mathbf{J}_1(r,\theta,z)$ is zero when $B = 0$, that is for $z = g_I/2(g_R^2 + g_I^2)$ the radial component disappears, then the transversal component of the Poynting vector is purely azimuthal. Furthermore, we can note that this effect is independent of the value of $n$; moreover, this effect does not dependent of the expression of the entire function $f$. As illustrative examples, we study the physical properties of Cauchy-Riemann beams determined by Eqs.\,(\ref{f_1})-(\ref{f_2}) with $n = 1, 3$. For the experimental generation of the fields, we employed a 4-$f$ optical setup. In this setup, the light beam from a He–Ne laser source is expanded by a spatial filter and collimated by a lens, this expanded beam illuminates a phase-only spatial light modulator (SLM), which displays a synthetic phase hologram (SPH). The SPH \cite{Arrizon07} consists of several terms, called signal and noise terms, so using a lens we apply its Fourier transform to separate the signal term, which is filtered and inverse Fourier transformed by a second lens to record the encoded field intensity on a CCD. Figs.\,\ref{fig_1}($\mathrm{a}_1$)-($\mathrm{a}_6$) show the experimental intensity pattern given by Eq.\,(\ref{psiCR}) considering Eq.\,(\ref{f_1}) with $n = 1,\,3$, at the planes $z = 0\,\mathrm{m}$, $z = z_0$ and $z = 2z_0$ where $z_0 = 2.57597\,\mathrm{m}$; with $z_0$ the value at which the radial component of the Poynting vector is zero. On the other hand, Figs.\,\ref{fig_1}($\mathrm{b}_1$)-($\mathrm{b}_6$) show the experimental intensity pattern given by Eq.\,(\ref{psiCR}) considering Eq.\,(\ref{f_2}) with $n = 1,\,3$, at the same planes $z$. In these plots, $g_R = .8 \times 10^{6}\,\mathrm{m}^{-2}$, $g_I = 1.5 \times 10^{6}\,\mathrm{m}^{-2}$, $\alpha = 8 \times 10^{4}\,\mathrm{m}$ and $\lambda = 633\,\mathrm{nm}$. From these plots, it is important to highlight the following, the Cauchy-Riemann beam rotates while it propagates, as can be observed in\,\cite{MoyaCauchy}. However, when the parameter $g$ is a complex number, the imaginary part of $g$ produces a focus of the intensity distribution, just up to the distance $z_0$, while for $z>z_0$, the effect is reverted, noting that the intensity pattern at $z = 0$ and $z = 2z_0$ is the same. Since the Cauchy-Riemann beam obtained by Eq.\,(\ref{f_1}) has  radial symmetry, the rotation effect can not be appreciable; nevertheless, for the Cauchy-Riemann beam determined by Eq.\,(\ref{f_2}) this effect can be noted. Although, the intensity pattern gives the information about the energy distribution, this is not enough. A more detailed description can be obtained from the phase profile and the vector Poynting, with which, we can determine the existence of optical vortices, as well as the energy flow given by the integral curves of the Poynting vector to determine if these Cauchy-Riemann beams can transfer angular momentum orbital. Finally, we comment that from the phase distribution and the Poynting vector it is possible to determine the vortices, since the integral curves of the Poynting vector (energy flow lines) describe concentric circles around these points. 

\begin{figure}
    \centering
    \includegraphics[width = \linewidth]{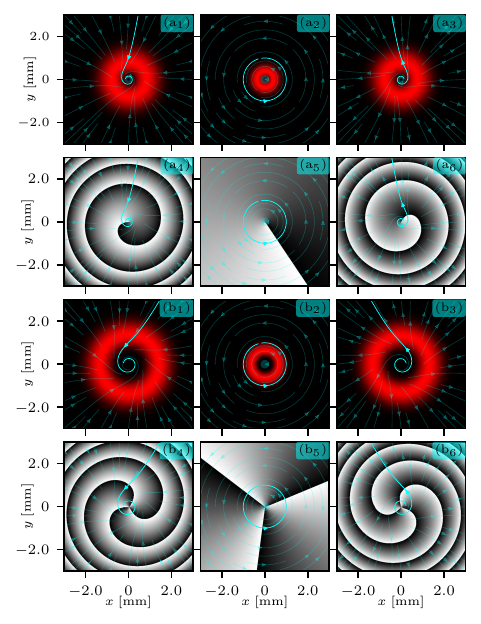}
    \caption{Intensity and phase distributions ($\mathrm{a}_1$)-($\mathrm{a}_3$) and ($\mathrm{a}_4$)-($\mathrm{a}_6$) determined by Eq.\,(\ref{psiCR}) considering $f_1\left(\frac{r}{r_\omega} e^{i(\theta-\theta_\omega)}\right)$ for $n = 1$ at the planes $z = 0\,\mathrm{m},\,z_0,\,2z_0$. Intensity and phase distributions ($\mathrm{b}_1$)-($\mathrm{b}_3$) and ($\mathrm{b}_4$)-($\mathrm{b}_6$) for $n = 3$ at the same planes of $z$. The cyan lines correspond to the energy flow lines. In these plots $g_R = .8 \times 10^{6}\,\mathrm{m}^{-2}$, $g_I = 1.5 \times 10^{6}\,\mathrm{m}^{-2}$, $\alpha = 8 \times 10^{4}\,\mathrm{m}$ and $\lambda = 633\,\mathrm{nm}$. The $x$ and $y$ coordinates ranging from  $-3\,\mathrm{mm}$ to $3\,\mathrm{mm}$.}
    \label{fig_2}
\end{figure}

\begin{figure}
    \centering
    \includegraphics[width = \linewidth]{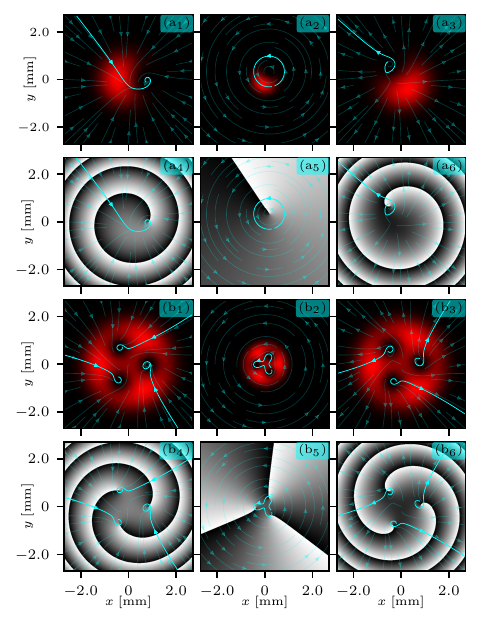}
    \caption{Intensity and phase distributions ($\mathrm{a}_1$)-($\mathrm{a}_3$) and ($\mathrm{a}_4$)-($\mathrm{a}_6$) determined by Eq.\,(\ref{psiCR}) considering $f_2\left(\frac{r}{r_\omega} e^{i(\theta-\theta_\omega)}\right)$ for $n = 1$ at the planes $z = 0\,\mathrm{m},\,z_0,\,2z_0$. Intensity and phase distributions ($\mathrm{b}_1$)-($\mathrm{b}_3$) and ($\mathrm{b}_4$)-($\mathrm{b}_6$) for $n = 3$ at the same planes of $z$. The cyan lines correspond to the energy flow lines. In these plots $g_R = .8 \times 10^{6}\,\mathrm{m}^{-2}$, $g_I = 1.5 \times 10^{6}\,\mathrm{m}^{-2}$, $\alpha = 8 \times 10^{4}\,\mathrm{m}$ and $\lambda = 633\,\mathrm{nm}$. The $x$ and $y$ coordinates ranging from  $-3\,\mathrm{mm}$ to $3\,\mathrm{mm}$.}
    \label{fig_3}
\end{figure}

In order to have a more detailed description on the Cauchy-Riemann beams, in Figs.\,\ref{fig_2}-\ref{fig_3} are shown the intensity and phase profiles, as well as the transversal component of the Poynting vector at the some representative planes of $z$ for the illustrative examples aforementioned. Figs.\,\ref{fig_2}($\mathrm{a}_1$)-($\mathrm{a}_6$) show the intensity and phase distributions, including the transversal energy flow lines, for the Cauchy-Riemann beam Eq.\,(\ref{psiCR}) considering Eq.\,(\ref{f_1}), for $n = 1$ at the planes $z = 0\,\mathrm{m}$, $z = z_0$ and $z = 2z_0$. Observe that, for $n = 1$ at the plane $z = 0\,\mathrm{m}$, the energy flow lines are directed towards the center; at the plane $z = z_0$ those have only azimuthal component, since the radial component vanishes; while at the plane $z = 2z_0$, the transversal energy flow lines changes in opposite direction at the plane $z = 0\,\mathrm{m}$. In general, in the interval $[0,z_0)$, this component tends to point towards the center. However, in the interval $(z_0,2 z_0]$, the direction of the transverse component reverses, pointing outward. Figs.\,\ref{fig_2}($\mathrm{b}_1$)-($\mathrm{c}_6$) are presented the corresponding plots of the intensity and phase distributions, likewise the transversal energy flow lines for the Cauchy-Riemann beam Eq.\,(\ref{psiCR}) considering Eq.\,(\ref{f_1}) for $n = 3$. The intensity pattern and the transversal energy flow lines have a similar behavior as the aforementioned case. Note that, the Cauchy-Riemann beams determined by Eqs.\,(\ref{f_1}), for $n = 1$ and $n = 3$, have an optical vortice at the point $(0,0)$ at plane $x-y$. The difference between both beams lies in the topological charge of their optical vortice. For $n = 1$, the optical vortice has topological charge equal to one; while for $n = 3$, the topological charge of the optical vortice is equal to three. Figs.\,\ref{fig_3}($\mathrm{a}_1$)-($\mathrm{a}_6$) show the intensity and phase distributions, in addition transversal component of the Poynting vector of the Cauchy-Riemann beam Eq.\,(\ref{psiCR}) taking into account Eq.\,(\ref{f_2}), for $n = 1$. As long as in Figs.\,\ref{fig_3}($\mathrm{b}_1$)-($\mathrm{b}_6$) are presented the corresponding plots for the Cauchy-Riemann beam Eq.\,(\ref{psiCR}) considering Eq.\,(\ref{f_2}) when $n = 3$. Both solutions have one and three optical vortices of the topological charge equal to one, respectively; however, the optical vortices of the Cauchy-Riemann beams Eq.\,(\ref{psiCR}) considering Eq.\,(\ref{f_2}) for $n = 1$ and $n = 3$ are not fixed points at the plane $x-y$. Moreover, these points follow a curve trajectory in the three-dimensional space.
 
\textit{Conclusions.} We have presented a detailed derivation of the Poynting vector for Cauchy-Riemann beams propagating in free space. The resulting analytical expressions provide insights into the energy flow and propagation dynamics of these beams. We also investigate Gaussian modulation with $g \in \mathbb{C}$, which can reverse the direction of energy flux and eliminate the radial component, resulting in a purely azimuthal field. Additionally, we validate our analytical results through experimental verification. Finally, we highlight that due to the behavior of the Poynting vector of  the Cauchy-Riemann beams studied in this letter, these beams can be used in different applications, such as, optical traps and transfer orbital angular momentum. 


\begin{thebibliography}{26}%
\makeatletter
\providecommand \@ifxundefined [1]{%
 \@ifx{#1\undefined}
}%
\providecommand \@ifnum [1]{%
 \ifnum #1\expandafter \@firstoftwo
 \else \expandafter \@secondoftwo
 \fi
}%
\providecommand \@ifx [1]{%
 \ifx #1\expandafter \@firstoftwo
 \else \expandafter \@secondoftwo
 \fi
}%
\providecommand \natexlab [1]{#1}%
\providecommand \enquote  [1]{``#1''}%
\providecommand \bibnamefont  [1]{#1}%
\providecommand \bibfnamefont [1]{#1}%
\providecommand \citenamefont [1]{#1}%
\providecommand \href@noop [0]{\@secondoftwo}%
\providecommand \href [0]{\begingroup \@sanitize@url \@href}%
\providecommand \@href[1]{\@@startlink{#1}\@@href}%
\providecommand \@@href[1]{\endgroup#1\@@endlink}%
\providecommand \@sanitize@url [0]{\catcode `\\12\catcode `\$12\catcode
  `\&12\catcode `\#12\catcode `\^12\catcode `\_12\catcode `\%12\relax}%
\providecommand \@@startlink[1]{}%
\providecommand \@@endlink[0]{}%
\providecommand \url  [0]{\begingroup\@sanitize@url \@url }%
\providecommand \@url [1]{\endgroup\@href {#1}{\urlprefix }}%
\providecommand \urlprefix  [0]{URL }%
\providecommand \Eprint [0]{\href }%
\providecommand \doibase [0]{https://doi.org/}%
\providecommand \selectlanguage [0]{\@gobble}%
\providecommand \bibinfo  [0]{\@secondoftwo}%
\providecommand \bibfield  [0]{\@secondoftwo}%
\providecommand \translation [1]{[#1]}%
\providecommand \BibitemOpen [0]{}%
\providecommand \bibitemStop [0]{}%
\providecommand \bibitemNoStop [0]{.\EOS\space}%
\providecommand \EOS [0]{\spacefactor3000\relax}%
\providecommand \BibitemShut  [1]{\csname bibitem#1\endcsname}%
\let\auto@bib@innerbib\@empty
\bibitem [{\citenamefont {Boyd}\ and\ \citenamefont
  {Gordon}(1961)}]{BoydHermite}%
  \BibitemOpen
  \bibfield  {author} {\bibinfo {author} {\bibfnamefont {G.~D.}\ \bibnamefont
  {Boyd}}\ and\ \bibinfo {author} {\bibfnamefont {J.~P.}\ \bibnamefont
  {Gordon}},\ }\bibfield  {title} {\bibinfo {title} {Confocal multimode
  resonator for millimeter through optical wavelength masers},\ }\href
  {https://doi.org/10.1002/j.1538-7305.1961.tb01626.x} {\bibfield  {journal}
  {\bibinfo  {journal} {Bell System Technical Journal}\ }\textbf {\bibinfo
  {volume} {40}},\ \bibinfo {pages} {489–508} (\bibinfo {year}
  {1961})}\BibitemShut {NoStop}%
\bibitem [{\citenamefont {Gori}\ \emph {et~al.}(1987)\citenamefont {Gori},
  \citenamefont {Guattari},\ and\ \citenamefont {Padovani}}]{GoriBesselGauss}%
  \BibitemOpen
  \bibfield  {author} {\bibinfo {author} {\bibfnamefont {F.}~\bibnamefont
  {Gori}}, \bibinfo {author} {\bibfnamefont {G.}~\bibnamefont {Guattari}},\
  and\ \bibinfo {author} {\bibfnamefont {C.}~\bibnamefont {Padovani}},\
  }\bibfield  {title} {\bibinfo {title} {Bessel-{G}auss beams},\ }\href
  {https://doi.org/https://doi.org/10.1016/0030-4018(87)90276-8} {\bibfield
  {journal} {\bibinfo  {journal} {Optics Communications}\ }\textbf {\bibinfo
  {volume} {64}},\ \bibinfo {pages} {491} (\bibinfo {year} {1987})}\BibitemShut
  {NoStop}%
\bibitem [{\citenamefont {Allen}\ \emph {et~al.}(1992)\citenamefont {Allen},
  \citenamefont {Beijersbergen}, \citenamefont {Spreeuw},\ and\ \citenamefont
  {Woerdman}}]{AllenOrbital}%
  \BibitemOpen
  \bibfield  {author} {\bibinfo {author} {\bibfnamefont {L.}~\bibnamefont
  {Allen}}, \bibinfo {author} {\bibfnamefont {M.~W.}\ \bibnamefont
  {Beijersbergen}}, \bibinfo {author} {\bibfnamefont {R.~J.~C.}\ \bibnamefont
  {Spreeuw}},\ and\ \bibinfo {author} {\bibfnamefont {J.~P.}\ \bibnamefont
  {Woerdman}},\ }\bibfield  {title} {\bibinfo {title} {Orbital angular momentum
  of light and the transformation of {L}aguerre-{G}aussian laser modes},\
  }\href {https://doi.org/10.1103/PhysRevA.45.8185} {\bibfield  {journal}
  {\bibinfo  {journal} {Phys. Rev. A}\ }\textbf {\bibinfo {volume} {45}},\
  \bibinfo {pages} {8185} (\bibinfo {year} {1992})}\BibitemShut {NoStop}%
\bibitem [{\citenamefont {Bandres}\ and\ \citenamefont
  {Guti\'{e}rrez-Vega}(2004)}]{BandresInce}%
  \BibitemOpen
  \bibfield  {author} {\bibinfo {author} {\bibfnamefont {M.~A.}\ \bibnamefont
  {Bandres}}\ and\ \bibinfo {author} {\bibfnamefont {J.~C.}\ \bibnamefont
  {Guti\'{e}rrez-Vega}},\ }\bibfield  {title} {\bibinfo {title}
  {Ince-{G}aussian beams},\ }\href {https://doi.org/10.1364/OL.29.000144}
  {\bibfield  {journal} {\bibinfo  {journal} {Opt. Lett.}\ }\textbf {\bibinfo
  {volume} {29}},\ \bibinfo {pages} {144} (\bibinfo {year} {2004})}\BibitemShut
  {NoStop}%
\bibitem [{\citenamefont {Siviloglou}\ and\ \citenamefont
  {Christodoulides}(2007)}]{SiviloglouAiry}%
  \BibitemOpen
  \bibfield  {author} {\bibinfo {author} {\bibfnamefont {G.~A.}\ \bibnamefont
  {Siviloglou}}\ and\ \bibinfo {author} {\bibfnamefont {D.~N.}\ \bibnamefont
  {Christodoulides}},\ }\bibfield  {title} {\bibinfo {title} {Accelerating
  finite energy {A}iry beams},\ }\href {https://doi.org/10.1364/OL.32.000979}
  {\bibfield  {journal} {\bibinfo  {journal} {Opt. Lett.}\ }\textbf {\bibinfo
  {volume} {32}},\ \bibinfo {pages} {979} (\bibinfo {year} {2007})}\BibitemShut
  {NoStop}%
\bibitem [{\citenamefont {Guti\'{e}rrez-Vega}\ and\ \citenamefont
  {Bandres}(2005)}]{GutierrezHelmholtz}%
  \BibitemOpen
  \bibfield  {author} {\bibinfo {author} {\bibfnamefont {J.~C.}\ \bibnamefont
  {Guti\'{e}rrez-Vega}}\ and\ \bibinfo {author} {\bibfnamefont {M.~A.}\
  \bibnamefont {Bandres}},\ }\bibfield  {title} {\bibinfo {title}
  {Helmholtz-{G}auss waves},\ }\href {https://doi.org/10.1364/JOSAA.22.000289}
  {\bibfield  {journal} {\bibinfo  {journal} {J. Opt. Soc. Am. A}\ }\textbf
  {\bibinfo {volume} {22}},\ \bibinfo {pages} {289} (\bibinfo {year}
  {2005})}\BibitemShut {NoStop}%
\bibitem [{\citenamefont {Lopez-Mago}\ \emph {et~al.}(2009)\citenamefont
  {Lopez-Mago}, \citenamefont {Bandres},\ and\ \citenamefont
  {Gutiérrez-Vega}}]{MagoPropagation}%
  \BibitemOpen
  \bibfield  {author} {\bibinfo {author} {\bibfnamefont {D.}~\bibnamefont
  {Lopez-Mago}}, \bibinfo {author} {\bibfnamefont {M.~A.}\ \bibnamefont
  {Bandres}},\ and\ \bibinfo {author} {\bibfnamefont {J.~C.}\ \bibnamefont
  {Gutiérrez-Vega}},\ }\bibfield  {title} {\bibinfo {title} {Propagation of
  {W}hittaker-{G}aussian beams},\ }in\ \href
  {https://doi.org/10.1117/12.825282} {\emph {\bibinfo {booktitle} {Laser Beam
  Shaping X}}},\ \bibinfo {editor} {edited by\ \bibinfo {editor} {\bibfnamefont
  {A.}~\bibnamefont {Forbes}}\ and\ \bibinfo {editor} {\bibfnamefont {T.~E.}\
  \bibnamefont {Lizotte}}}\ (\bibinfo  {publisher} {SPIE},\ \bibinfo {year}
  {2009})\BibitemShut {NoStop}%
\bibitem [{\citenamefont {Esp\'{i}ndola-Ramos}\ \emph
  {et~al.}(2019)\citenamefont {Esp\'{i}ndola-Ramos}, \citenamefont
  {Silva-Ortigoza}, \citenamefont {Sosa-S\'{a}nchez}, \citenamefont
  {Juli\'{a}n-Mac\'{i}as}, \citenamefont {de~J.~Cabrera-Rosas}, \citenamefont
  {Ortega-Vidals}, \citenamefont {Gonz\'{a}lez-Ju\'{a}rez}, \citenamefont
  {Silva-Ortigoza}, \citenamefont {Vel\'{a}zquez-Quesada},\ and\ \citenamefont
  {del Castillo}}]{EspindolaParaxial}%
  \BibitemOpen
  \bibfield  {author} {\bibinfo {author} {\bibfnamefont {E.}~\bibnamefont
  {Esp\'{i}ndola-Ramos}}, \bibinfo {author} {\bibfnamefont {G.}~\bibnamefont
  {Silva-Ortigoza}}, \bibinfo {author} {\bibfnamefont {C.~T.}\ \bibnamefont
  {Sosa-S\'{a}nchez}}, \bibinfo {author} {\bibfnamefont {I.}~\bibnamefont
  {Juli\'{a}n-Mac\'{i}as}}, \bibinfo {author} {\bibfnamefont {O.}~\bibnamefont
  {de~J.~Cabrera-Rosas}}, \bibinfo {author} {\bibfnamefont {P.}~\bibnamefont
  {Ortega-Vidals}}, \bibinfo {author} {\bibfnamefont {A.}~\bibnamefont
  {Gonz\'{a}lez-Ju\'{a}rez}}, \bibinfo {author} {\bibfnamefont
  {R.}~\bibnamefont {Silva-Ortigoza}}, \bibinfo {author} {\bibfnamefont
  {M.~P.}\ \bibnamefont {Vel\'{a}zquez-Quesada}},\ and\ \bibinfo {author}
  {\bibfnamefont {G.~F.~T.}\ \bibnamefont {del Castillo}},\ }\bibfield  {title}
  {\bibinfo {title} {Paraxial optical fields whose intensity pattern skeletons
  are stable caustics},\ }\href {https://doi.org/10.1364/JOSAA.36.001820}
  {\bibfield  {journal} {\bibinfo  {journal} {J. Opt. Soc. Am. A}\ }\textbf
  {\bibinfo {volume} {36}},\ \bibinfo {pages} {1820} (\bibinfo {year}
  {2019})}\BibitemShut {NoStop}%
\bibitem [{\citenamefont {Julián-Macías}\ \emph {et~al.}(2020)\citenamefont
  {Julián-Macías}, \citenamefont {Sosa-Sánchez}, \citenamefont
  {de~J.~Cabrera-Rosas}, \citenamefont {Espíndola-Ramos},\ and\ \citenamefont
  {Silva-Ortigoza}}]{JulianThevector}%
  \BibitemOpen
  \bibfield  {author} {\bibinfo {author} {\bibfnamefont {I.}~\bibnamefont
  {Julián-Macías}}, \bibinfo {author} {\bibfnamefont {C.~T.}\ \bibnamefont
  {Sosa-Sánchez}}, \bibinfo {author} {\bibfnamefont {O.}~\bibnamefont
  {de~J.~Cabrera-Rosas}}, \bibinfo {author} {\bibfnamefont {E.}~\bibnamefont
  {Espíndola-Ramos}},\ and\ \bibinfo {author} {\bibfnamefont {G.}~\bibnamefont
  {Silva-Ortigoza}},\ }\bibfield  {title} {\bibinfo {title} {The vector
  {D}urnin–{W}hitney beam},\ }\href {https://doi.org/10.1364/josaa.376545}
  {\bibfield  {journal} {\bibinfo  {journal} {Journal of the Optical Society of
  America A}\ }\textbf {\bibinfo {volume} {37}},\ \bibinfo {pages} {294}
  (\bibinfo {year} {2020})}\BibitemShut {NoStop}%
\bibitem [{\citenamefont {Singh}\ \emph {et~al.}(2023)\citenamefont {Singh},
  \citenamefont {Kinashi}, \citenamefont {Tsutsumi}, \citenamefont {Sakai},\
  and\ \citenamefont {Jackin}}]{SinghTricomiGauss}%
  \BibitemOpen
  \bibfield  {author} {\bibinfo {author} {\bibfnamefont {S.~K.}\ \bibnamefont
  {Singh}}, \bibinfo {author} {\bibfnamefont {K.}~\bibnamefont {Kinashi}},
  \bibinfo {author} {\bibfnamefont {N.}~\bibnamefont {Tsutsumi}}, \bibinfo
  {author} {\bibfnamefont {W.}~\bibnamefont {Sakai}},\ and\ \bibinfo {author}
  {\bibfnamefont {B.~J.}\ \bibnamefont {Jackin}},\ }\bibfield  {title}
  {\bibinfo {title} {Tricomi–{G}auss beam and its propagation
  characteristics},\ }\bibfield  {journal} {\bibinfo  {journal} {Optical and
  Quantum Electronics}\ }\textbf {\bibinfo {volume} {55}},\ \href
  {https://doi.org/10.1007/s11082-023-04626-x} {10.1007/s11082-023-04626-x}
  (\bibinfo {year} {2023})\BibitemShut {NoStop}%
\bibitem [{\citenamefont {Moya-Cessa}\ \emph {et~al.}(2024)\citenamefont
  {Moya-Cessa}, \citenamefont {Ramos-Prieto}, \citenamefont {S\'anchez-de-la
  Llave}, \citenamefont {Ru\'{\i}z}, \citenamefont {Arriz\'on},\ and\
  \citenamefont {Soto-Eguibar}}]{MoyaCauchy}%
  \BibitemOpen
  \bibfield  {author} {\bibinfo {author} {\bibfnamefont {H.~M.}\ \bibnamefont
  {Moya-Cessa}}, \bibinfo {author} {\bibfnamefont {I.}~\bibnamefont
  {Ramos-Prieto}}, \bibinfo {author} {\bibfnamefont {D.}~\bibnamefont
  {S\'anchez-de-la Llave}}, \bibinfo {author} {\bibfnamefont {U.}~\bibnamefont
  {Ru\'{\i}z}}, \bibinfo {author} {\bibfnamefont {V.}~\bibnamefont
  {Arriz\'on}},\ and\ \bibinfo {author} {\bibfnamefont {F.}~\bibnamefont
  {Soto-Eguibar}},\ }\bibfield  {title} {\bibinfo {title} {Cauchy-{R}iemann
  beams},\ }\href {https://doi.org/10.1103/PhysRevA.109.043528} {\bibfield
  {journal} {\bibinfo  {journal} {Phys. Rev. A}\ }\textbf {\bibinfo {volume}
  {109}},\ \bibinfo {pages} {043528} (\bibinfo {year} {2024})}\BibitemShut
  {NoStop}%
\bibitem [{\citenamefont {Volke-Sep\'{u}lveda}\ \emph
  {et~al.}(2004)\citenamefont {Volke-Sep\'{u}lveda}, \citenamefont
  {Ch\'{a}vez-Cerda}, \citenamefont {Garc\'{e}s-Ch\'{a}vez},\ and\
  \citenamefont {Dholakia}}]{VolkeThreedimensional}%
  \BibitemOpen
  \bibfield  {author} {\bibinfo {author} {\bibfnamefont {K.}~\bibnamefont
  {Volke-Sep\'{u}lveda}}, \bibinfo {author} {\bibfnamefont {S.}~\bibnamefont
  {Ch\'{a}vez-Cerda}}, \bibinfo {author} {\bibfnamefont {V.}~\bibnamefont
  {Garc\'{e}s-Ch\'{a}vez}},\ and\ \bibinfo {author} {\bibfnamefont
  {K.}~\bibnamefont {Dholakia}},\ }\bibfield  {title} {\bibinfo {title}
  {Three-dimensional optical forces and transfer of orbital angular momentum
  from multiringed light beams to spherical microparticles},\ }\href
  {https://doi.org/10.1364/JOSAB.21.001749} {\bibfield  {journal} {\bibinfo
  {journal} {J. Opt. Soc. Am. B}\ }\textbf {\bibinfo {volume} {21}},\ \bibinfo
  {pages} {1749} (\bibinfo {year} {2004})}\BibitemShut {NoStop}%
\bibitem [{\citenamefont {Padgett}\ and\ \citenamefont
  {Allen}(1995)}]{PadgetThePoynitng}%
  \BibitemOpen
  \bibfield  {author} {\bibinfo {author} {\bibfnamefont {M.}~\bibnamefont
  {Padgett}}\ and\ \bibinfo {author} {\bibfnamefont {L.}~\bibnamefont
  {Allen}},\ }\bibfield  {title} {\bibinfo {title} {The {P}oynting vector in
  {L}aguerre-{G}aussian laser modes},\ }\href
  {https://doi.org/https://doi.org/10.1016/0030-4018(95)00455-H} {\bibfield
  {journal} {\bibinfo  {journal} {Optics Communications}\ }\textbf {\bibinfo
  {volume} {121}},\ \bibinfo {pages} {36} (\bibinfo {year} {1995})}\BibitemShut
  {NoStop}%
\bibitem [{\citenamefont {Allen}\ \emph {et~al.}(1999)\citenamefont {Allen},
  \citenamefont {Padgett},\ and\ \citenamefont {Babiker}}]{AllenIVTheOrbital}%
  \BibitemOpen
  \bibfield  {author} {\bibinfo {author} {\bibfnamefont {L.}~\bibnamefont
  {Allen}}, \bibinfo {author} {\bibfnamefont {M.}~\bibnamefont {Padgett}},\
  and\ \bibinfo {author} {\bibfnamefont {M.}~\bibnamefont {Babiker}},\
  }\bibinfo {title} {Iv the orbital angular momentum of light},\ in\ \href
  {https://doi.org/10.1016/s0079-6638(08)70391-3} {\emph {\bibinfo {booktitle}
  {Progress in Optics}}}\ (\bibinfo  {publisher} {Elsevier},\ \bibinfo {year}
  {1999})\ p.\ \bibinfo {pages} {291–372}\BibitemShut {NoStop}%
\bibitem [{\citenamefont {Ostrovsky}\ \emph {et~al.}(2013)\citenamefont
  {Ostrovsky}, \citenamefont {Rickenstorff-Parrao},\ and\ \citenamefont
  {Arriz\'{o}n}}]{OstrovskyLiquid}%
  \BibitemOpen
  \bibfield  {author} {\bibinfo {author} {\bibfnamefont {A.~S.}\ \bibnamefont
  {Ostrovsky}}, \bibinfo {author} {\bibfnamefont {C.}~\bibnamefont
  {Rickenstorff-Parrao}},\ and\ \bibinfo {author} {\bibfnamefont
  {V.}~\bibnamefont {Arriz\'{o}n}},\ }\bibfield  {title} {\bibinfo {title}
  {Generation of the \&\#x201c;perfect\&\#x201d; optical vortex using a
  liquid-crystal spatial light modulator},\ }\href
  {https://doi.org/10.1364/OL.38.000534} {\bibfield  {journal} {\bibinfo
  {journal} {Opt. Lett.}\ }\textbf {\bibinfo {volume} {38}},\ \bibinfo {pages}
  {534} (\bibinfo {year} {2013})}\BibitemShut {NoStop}%
\bibitem [{\citenamefont {Chen}\ \emph {et~al.}(2013)\citenamefont {Chen},
  \citenamefont {Mazilu}, \citenamefont {Arita}, \citenamefont {Wright},\ and\
  \citenamefont {Dholakia}}]{ChenDynamics}%
  \BibitemOpen
  \bibfield  {author} {\bibinfo {author} {\bibfnamefont {M.}~\bibnamefont
  {Chen}}, \bibinfo {author} {\bibfnamefont {M.}~\bibnamefont {Mazilu}},
  \bibinfo {author} {\bibfnamefont {Y.}~\bibnamefont {Arita}}, \bibinfo
  {author} {\bibfnamefont {E.~M.}\ \bibnamefont {Wright}},\ and\ \bibinfo
  {author} {\bibfnamefont {K.}~\bibnamefont {Dholakia}},\ }\bibfield  {title}
  {\bibinfo {title} {Dynamics of microparticles trapped in a perfect vortex
  beam},\ }\href {https://doi.org/10.1364/OL.38.004919} {\bibfield  {journal}
  {\bibinfo  {journal} {Opt. Lett.}\ }\textbf {\bibinfo {volume} {38}},\
  \bibinfo {pages} {4919} (\bibinfo {year} {2013})}\BibitemShut {NoStop}%
\bibitem [{\citenamefont {Ramos-Prieto}\ \emph {et~al.}(2024)\citenamefont
  {Ramos-Prieto}, \citenamefont {de-la Llave}, \citenamefont {Ruíz},
  \citenamefont {Arriz\'on}, \citenamefont {Soto-Eguibar},\ and\ \citenamefont
  {Moya-Cessa}}]{IranCauchyGRIN}%
  \BibitemOpen
  \bibfield  {author} {\bibinfo {author} {\bibfnamefont {I.}~\bibnamefont
  {Ramos-Prieto}}, \bibinfo {author} {\bibfnamefont {D.~S.}\ \bibnamefont
  {de-la Llave}}, \bibinfo {author} {\bibfnamefont {U.}~\bibnamefont {Ruíz}},
  \bibinfo {author} {\bibfnamefont {V.}~\bibnamefont {Arriz\'on}}, \bibinfo
  {author} {\bibfnamefont {F.}~\bibnamefont {Soto-Eguibar}},\ and\ \bibinfo
  {author} {\bibfnamefont {H.}~\bibnamefont {Moya-Cessa}},\ }\bibfield  {title}
  {\bibinfo {title} {Cauchy–{R}iemann beams in {GRIN} media},\ }\href@noop {}
  {\bibfield  {journal} {\bibinfo  {journal} {Optik}\ }\textbf {\bibinfo
  {volume} {309}},\ \bibinfo {pages} {171864} (\bibinfo {year}
  {2024})}\BibitemShut {NoStop}%
\bibitem [{\citenamefont {Stoler}(1981)}]{Stoler}%
  \BibitemOpen
  \bibfield  {author} {\bibinfo {author} {\bibfnamefont {D.}~\bibnamefont
  {Stoler}},\ }\bibfield  {title} {\bibinfo {title} {Operator methods in
  physical optics},\ }\href {https://doi.org/10.1364/JOSA.71.000334} {\bibfield
   {journal} {\bibinfo  {journal} {J. Opt. Soc. Am.}\ }\textbf {\bibinfo
  {volume} {71}},\ \bibinfo {pages} {334} (\bibinfo {year} {1981})}\BibitemShut
  {NoStop}%
\bibitem [{\citenamefont {Korneev}\ \emph
  {et~al.}(2024{\natexlab{a}})\citenamefont {Korneev}, \citenamefont
  {Ramos-Prieto}, \citenamefont {Soto-Eguibar}, \citenamefont {Ruíz},
  \citenamefont {de-la Llave},\ and\ \citenamefont
  {Moya-Cessa}}]{KorneevUnified}%
  \BibitemOpen
  \bibfield  {author} {\bibinfo {author} {\bibfnamefont {N.}~\bibnamefont
  {Korneev}}, \bibinfo {author} {\bibfnamefont {I.}~\bibnamefont
  {Ramos-Prieto}}, \bibinfo {author} {\bibfnamefont {F.}~\bibnamefont
  {Soto-Eguibar}}, \bibinfo {author} {\bibfnamefont {U.}~\bibnamefont {Ruíz}},
  \bibinfo {author} {\bibfnamefont {D.~S.}\ \bibnamefont {de-la Llave}},\ and\
  \bibinfo {author} {\bibfnamefont {H.~M.}\ \bibnamefont {Moya-Cessa}},\ }\href
  {https://arxiv.org/abs/2405.20548} {\bibinfo {title} {Unified approach to
  paraxial propagation in uniform media and media with linear or quadratic
  refractive index distribution}} (\bibinfo {year} {2024}{\natexlab{a}}),\
  \Eprint {https://arxiv.org/abs/2405.20548} {arXiv:2405.20548
  [physics.optics]} \BibitemShut {NoStop}%
\bibitem [{\citenamefont {Korneev}\ \emph
  {et~al.}(2024{\natexlab{b}})\citenamefont {Korneev}, \citenamefont
  {Ramos-Prieto}, \citenamefont {Julián-Macías}, \citenamefont {Ruíz},
  \citenamefont {Soto-Eguibar}, \citenamefont {de-la Llave},\ and\
  \citenamefont {Moya-Cessa}}]{KorneevAsymmetric}%
  \BibitemOpen
  \bibfield  {author} {\bibinfo {author} {\bibfnamefont {N.}~\bibnamefont
  {Korneev}}, \bibinfo {author} {\bibfnamefont {I.}~\bibnamefont
  {Ramos-Prieto}}, \bibinfo {author} {\bibfnamefont {I.}~\bibnamefont
  {Julián-Macías}}, \bibinfo {author} {\bibfnamefont {U.}~\bibnamefont
  {Ruíz}}, \bibinfo {author} {\bibfnamefont {F.}~\bibnamefont {Soto-Eguibar}},
  \bibinfo {author} {\bibfnamefont {D.~S.}\ \bibnamefont {de-la Llave}},\ and\
  \bibinfo {author} {\bibfnamefont {H.~M.}\ \bibnamefont {Moya-Cessa}},\ }\href
  {https://arxiv.org/abs/2408.11953} {\bibinfo {title} {Asymmetric
  {C}auchy-{R}iemann beams}} (\bibinfo {year} {2024}{\natexlab{b}}),\ \Eprint
  {https://arxiv.org/abs/2408.11953} {arXiv:2408.11953 [physics.optics]}
  \BibitemShut {NoStop}%
\bibitem [{\citenamefont {Schechner}\ \emph {et~al.}(1996)\citenamefont
  {Schechner}, \citenamefont {Piestun},\ and\ \citenamefont
  {Shamir}}]{SchechnerWave}%
  \BibitemOpen
  \bibfield  {author} {\bibinfo {author} {\bibfnamefont {Y.~Y.}\ \bibnamefont
  {Schechner}}, \bibinfo {author} {\bibfnamefont {R.}~\bibnamefont {Piestun}},\
  and\ \bibinfo {author} {\bibfnamefont {J.}~\bibnamefont {Shamir}},\
  }\bibfield  {title} {\bibinfo {title} {Wave propagation with rotating
  intensity distributions},\ }\href {https://doi.org/10.1103/physreve.54.r50}
  {\bibfield  {journal} {\bibinfo  {journal} {Physical Review E}\ }\textbf
  {\bibinfo {volume} {54}},\ \bibinfo {pages} {R50–R53} (\bibinfo {year}
  {1996})}\BibitemShut {NoStop}%
\bibitem [{\citenamefont {Bekshaev}\ \emph {et~al.}(2004)\citenamefont
  {Bekshaev}, \citenamefont {Soskin},\ and\ \citenamefont
  {Vasnetsov}}]{BekshaevAnoptical}%
  \BibitemOpen
  \bibfield  {author} {\bibinfo {author} {\bibfnamefont {A.~Y.}\ \bibnamefont
  {Bekshaev}}, \bibinfo {author} {\bibfnamefont {M.~S.}\ \bibnamefont
  {Soskin}},\ and\ \bibinfo {author} {\bibfnamefont {M.~V.}\ \bibnamefont
  {Vasnetsov}},\ }\bibfield  {title} {\bibinfo {title} {An optical vortex as a
  rotating body: mechanical features of a singular light beam},\ }\href
  {https://doi.org/10.1088/1464-4258/6/5/004} {\bibfield  {journal} {\bibinfo
  {journal} {Journal of Optics A: Pure and Applied Optics}\ }\textbf {\bibinfo
  {volume} {6}},\ \bibinfo {pages} {S170–S174} (\bibinfo {year}
  {2004})}\BibitemShut {NoStop}%
\bibitem [{\citenamefont {Razueva}\ and\ \citenamefont
  {Abramochkin}(2019)}]{RazuevaMultiple}%
  \BibitemOpen
  \bibfield  {author} {\bibinfo {author} {\bibfnamefont {E.}~\bibnamefont
  {Razueva}}\ and\ \bibinfo {author} {\bibfnamefont {E.}~\bibnamefont
  {Abramochkin}},\ }\bibfield  {title} {\bibinfo {title} {Multiple-twisted
  spiral beams},\ }\href {https://doi.org/10.1364/josaa.36.001089} {\bibfield
  {journal} {\bibinfo  {journal} {Journal of the Optical Society of America A}\
  }\textbf {\bibinfo {volume} {36}},\ \bibinfo {pages} {1089} (\bibinfo {year}
  {2019})}\BibitemShut {NoStop}%
\bibitem [{\citenamefont {Dummit}\ and\ \citenamefont
  {{F}oote}(2004)}]{Dummit}%
  \BibitemOpen
  \bibfield  {author} {\bibinfo {author} {\bibfnamefont {D.~S.}\ \bibnamefont
  {Dummit}}\ and\ \bibinfo {author} {\bibfnamefont {R.~M.}\ \bibnamefont
  {{F}oote}},\ }\href@noop {} {\emph {\bibinfo {title} {Abstract Algebra}}},\
  Vol.~\bibinfo {volume} {3}\ (\bibinfo  {publisher} {Wiley Hoboken},\ \bibinfo
  {year} {2004})\BibitemShut {NoStop}%
\bibitem [{\citenamefont {Berry}\ and\ \citenamefont
  {McDonald}(2008)}]{BerryExact}%
  \BibitemOpen
  \bibfield  {author} {\bibinfo {author} {\bibfnamefont {M.~V.}\ \bibnamefont
  {Berry}}\ and\ \bibinfo {author} {\bibfnamefont {K.~T.}\ \bibnamefont
  {McDonald}},\ }\bibfield  {title} {\bibinfo {title} {Exact and geometrical
  optics energy trajectories in twisted beams},\ }\href
  {https://doi.org/10.1088/1464-4258/10/3/035005} {\bibfield  {journal}
  {\bibinfo  {journal} {Journal of Optics A: Pure and Applied Optics}\ }\textbf
  {\bibinfo {volume} {10}},\ \bibinfo {pages} {035005} (\bibinfo {year}
  {2008})}\BibitemShut {NoStop}%
\bibitem [{\citenamefont {Arriz\'{o}n}\ \emph {et~al.}(2007)\citenamefont
  {Arriz\'{o}n}, \citenamefont {Ruiz}, \citenamefont {Carrada},\ and\
  \citenamefont {Gonz\'{a}lez}}]{Arrizon07}%
  \BibitemOpen
  \bibfield  {author} {\bibinfo {author} {\bibfnamefont {V.}~\bibnamefont
  {Arriz\'{o}n}}, \bibinfo {author} {\bibfnamefont {U.}~\bibnamefont {Ruiz}},
  \bibinfo {author} {\bibfnamefont {R.}~\bibnamefont {Carrada}},\ and\ \bibinfo
  {author} {\bibfnamefont {L.~A.}\ \bibnamefont {Gonz\'{a}lez}},\ }\bibfield
  {title} {\bibinfo {title} {Pixelated phase computer holograms for the
  accurate encoding of scalar complex fields},\ }\href
  {https://doi.org/10.1364/JOSAA.24.003500} {\bibfield  {journal} {\bibinfo
  {journal} {J. Opt. Soc. Am. A}\ }\textbf {\bibinfo {volume} {24}},\ \bibinfo
  {pages} {3500} (\bibinfo {year} {2007})}\BibitemShut {NoStop}%
\end{thebibliography}

%

\end{document}